\documentclass{Interspeech}

\interspeechcameraready

\title{Mitigating Audiovisual Mismatch in Visual-Guide Audio Captioning}

\author[affiliation={1}]{Le}{Xu}
\author[affiliation={1}, equalcontribution]{Chenxing}{Li}
\author[affiliation={1,3}]{Yong}{Ren}
\author[affiliation={1,3}]{Yujie}{Chen}
\author[affiliation={1}]{Yu}{Gu}
\author[affiliation={3}]{Ruibo}{Fu}
\author[affiliation={1}]{Shan}{Yang}
\author[affiliation={2}, equalcontribution]{Dong}{Yu}

\affiliation{}{Tencent AI Lab}{Beijing, China}
\affiliation{}{Tencent AI Lab}{Seattle, USA}
\affiliation{}{Institute of Automation, Chinese Academy of Sciences}{Beijing, China}
\email{lichenxing007@gmail.com, dongyu@ieee.org}

\usepackage{comment}
\usepackage{multirow}
\usepackage{algorithm}
\usepackage{algorithmic}
\usepackage{graphicx}
\usepackage{subcaption}

\begin{document}

\maketitle

\begin{abstract}
    Current vision-guided audio captioning systems frequently fail to address audiovisual misalignment in real-world scenarios, such as dubbed content or off-screen sounds. To bridge this critical gap, we present an entropy-aware gated fusion framework that dynamically modulates visual information flow through cross-modal uncertainty quantification. Our novel approach employs attention entropy analysis in cross-attention layers to automatically identify and suppress misleading visual cues during modal fusion. Complementing this architecture, we develop a batch-wise audiovisual shuffling technique that generates synthetic mismatched training pairs, greatly enhancing model resilience against alignment noise. Evaluations on the AudioCaps benchmark demonstrate our system's superior performance over existing baselines, especially in mismatched modality scenarios. Furthermore, our solution demonstrates an approximately 6x improvement in inference speed compared to the baseline.
\end{abstract}

\section{Introduction}
    Automatic audio captioning (AAC), which aims to generate natural language descriptions for unconstrained audio clips \cite{mei2022automated}, has emerged as a critical technology bridging machine perception and human communication. This task is applicable in areas such as multimedia retrieval \cite{mei2022metric,liu2022separate}, assistive technologies for the hearing-impaired \cite{hong2010dynamic, hong2011video}, and intelligent video analysis systems \cite{lee2024nowyousee}. While existing approaches predominantly focus on modelling audio signals alone \cite{xu2024towards,mei2021act,drossos2020clotho, liu2024enhancing}, recent studies have explored leveraging visual information from accompanying videos to understand complex auditory scenes better \cite{liu2022vact, kim24g_avcap, rho2025lavcap}. However, these vision-guided approaches typically rely on the assumption of perfect audiovisual correspondence, an assumption that often does not hold in real-world scenarios due to issues like dubbed content or off-screen sound sources.

    Previous multimodal AAC frameworks generally employ early fusion strategies (e.g., concatenation-based feature merging \cite{kim24g_avcap,rho2025lavcap}) or late fusion architectures with attention mechanisms \cite{vaswani2017attention, liu2022vact} to combine audio features with visual representations extracted from synchronized video frames. AVCap \cite{kim24g_avcap} concatenates raw audiovisual features, assuming inherent compatibility between modalities. Building on this, LAVCap \cite{rho2025lavcap} introduces an optimal transport-based loss function to align audiovisual modalities better. VACT \cite{liu2022vact} proposes an Adaptive Audio-Visual Attention method that integrates audio and visual information through confidence scores derived from the textual modality. However, these approaches face two significant limitations: First, they assume semantic consistency between the modalities, which can introduce noise when this assumption is violated, such as when ambient sounds come from off-screen objects or when visible objects have no corresponding sounds. This prevalent but overlooked audiovisual mismatch problem can significantly degrade model performance by misleading cross-modal reasoning. Second, when processing long-form video inputs, concatenation-based systems incur prohibitive computational costs, as they require redundant encoding of dense frame-level visual features even for temporally static acoustic events. This computational inefficiency becomes particularly problematic in real-time applications with strict latency requirements.

To tackle this challenge, we introduce a novel framework that incorporates entropy-aware gated cross-attention with contrastive data augmentation. Our key insight lies in enabling dynamic modality selection rather than enforcing unconditional visual dependency. The proposed gate mechanism evaluates cross-modality relevance by calculating attention entropy during feature interactions, which allows for the automatic suppression of misleading visual cues during caption generation. Additionally, our mismatch-aware data augmentation strategy systematically exposes the model to mismatched audiovisual samples during training, which fosters robustness against real-world inconsistencies. The primary contributions of this work include: 
\begin{enumerate}
    \item First systematic investigation of audiovisual mismatch in video-assisted AAC, identifying its detrimental effects on conventional fusion approaches; 
    \item An adaptive gating mechanism based on attention entropy dynamics, achieving context-aware modality weighting without additional supervision; 
    \item A practical learning paradigm through randomized intra-batch re-pairing, effectively simulating diverse mismatch scenarios;
    \item Comprehensive evaluations across multiple metrics demonstrate superior performance under mismatched conditions. Moreover, our system achieves an approximately 6x improvement in inference speed compared to the baseline while still maintaining competitive accuracy.
\end{enumerate}


\begin{figure*}[thb]
    \centering
    \includegraphics[width=0.95\linewidth]{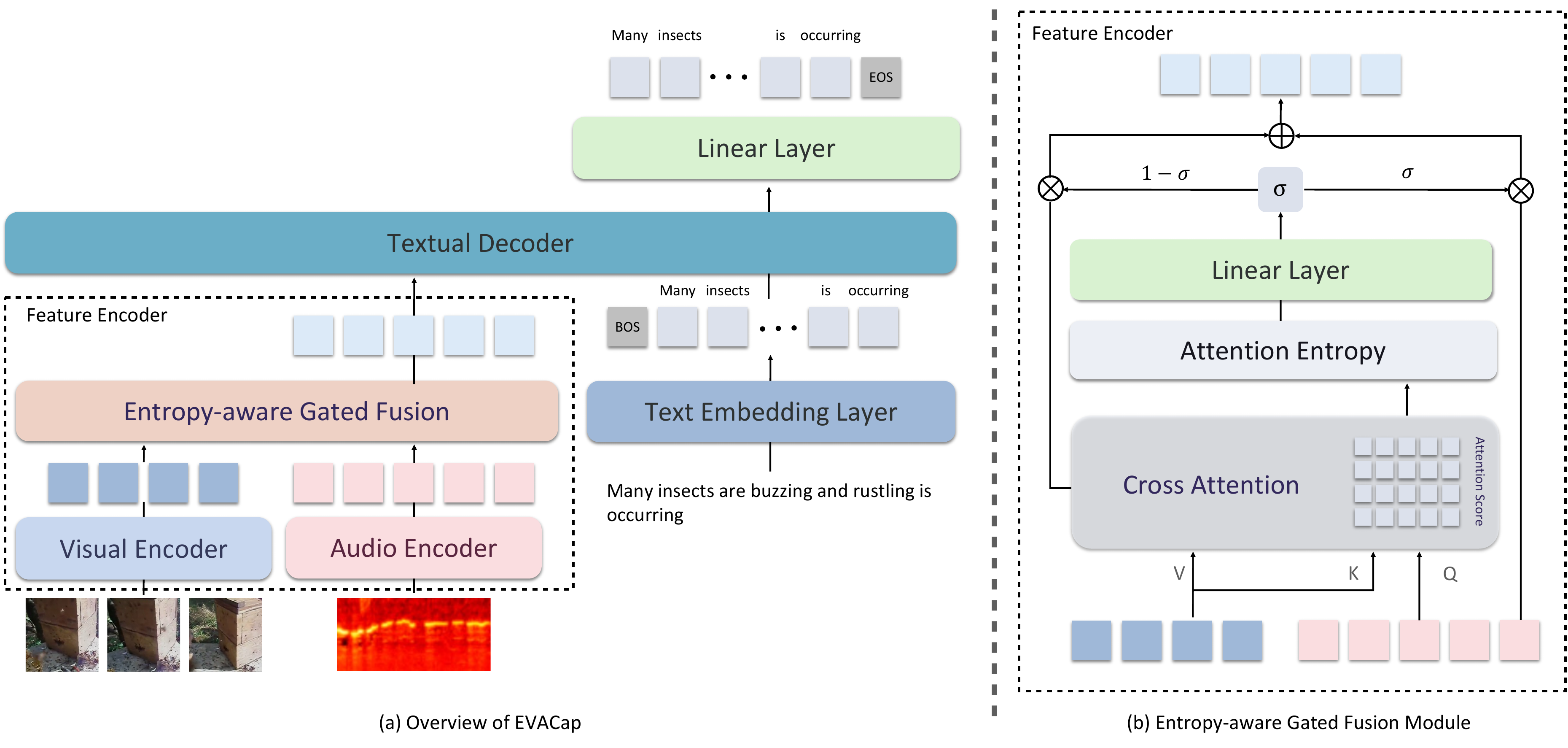}
    \caption{(a) Overview of the proposed EVACap Framework. (b) Detail of the Entropy-aware Gated Fusion module.}
    \label{fig:arch}
\end{figure*}

\section{Methods}

Our proposed framework, \textbf{E}ntropy-aware \textbf{V}isual-guide \textbf{A}udio \textbf{Cap}tioning (called \textbf{EVACap}), addresses audiovisual mismatches with two key innovations: (1) an entropy-guided gated fusion mechanism that dynamically controls the flow of visual information, and (2) a probabilistic data augmentation strategy that enhances model robustness. The overall architecture is based on the CAV-MAE \cite{gong2022cavmae} pretrained foundation, with careful adaptations tailored for caption generation.

\subsection{Audiovisual modal alignment}

As shown in Figure \ref{fig:arch}(a). We employ CAV-MAE \cite{gong2022cavmae} as our multimodal encoder due to its contrastive audiovisual alignment capabilities. For each input video clip ($a$,$v$), we extract temporal features from both modalities:

\begin{itemize}
\item Visual features $F_v \in \mathbb{R}^{T_v \times d}$ using frame-based Vision Transformers (ViT) \cite{dosovitskiy2020vit} encoder;
\item Audio features $F_a \in \mathbb{R}^{T_a \times d}$ via fbank processing with same struct of ViT encoder.
\end{itemize}

Unlike the original CAV-MAE modal fusion method, our novel entropy-gated fusion module facilitates conditional interaction. This approach preserves modality-specific characteristics while allowing for context-aware integration of visual cues.

\subsection{Entropy-aware gated modal fusion}
The core innovation lies in quantifying the relevance of the cross-modal relationship through the analysis of the distribution of attention. Given a query vector $Q$ from audio $A$ and key-value pairs ($K$,$V$) from vision, where $T_v$ and $T_a$ denote the length of the visual and audio input sequences respectively:

\begin{equation}
\begin{aligned}
    &P_{att} = \text{softmax}(\frac{QK^T}{\sqrt{d}}), \\
    &F = P_{att} \times V.
\end{aligned}
\end{equation}

We compute attention scores $P_{att}$ and multimodal features $F$. Then we obtain the entropy $E_{att}$ over the normalized weights $\alpha \in \mathbb{R^N}$ where $N = T_v$ and $p \in P_{att}$, measuring the dispersion of visual focus:

\begin{equation}
    E_{att} = \frac{
        - \sum^{T_a}_{i=1}p_i\log{p_i}
        }{
        \log{\alpha}
    }.
\end{equation}


As shown in Figure \ref{fig:arch} (b). A learnable gate modulates visual influence based on this uncertainty metric:


\begin{equation}
\begin{aligned}
    &g = \sigma(w_g E_{\text{att}} + b_g), \\
    &F = (1-g) \cdot A + g \cdot F,
\end{aligned}
\end{equation}
where $w_g$, $b_g$ are trainable parameters and $\sigma$ denotes sigmoid activation. The system utilizes the gating value $g$ to identify and suppress misleading visual cues during modal fusion automatically. 

\subsection{Mismatch-aware data augmentation}

To explicitly handle real-world audiovisual discrepancies, we propose Stochastic Modality Shuffling (SMS) during training. Let $\mathcal{B} = \{(a_i,v_i)\}^B_{i=1}$ denote a batch containing $B$ samples, and $R_B$ denote the set of all permutation matrices of size $B$:

\begin{algorithm}[hb]
\caption{Stochastic Modality Shuffling}
\begin{algorithmic}[1]
\STATE Input: Batch features $\mathbf{V} = [v_1,\dots,v_B]$
\STATE Generate permutation matrix $P \sim U(R_B)$

\STATE Apply shuffle: $\hat{\mathbf{V}} = P\mathbf{V}$
\STATE Mix with original audio: $\hat{\mathcal{B}} = \{(a_i, \hat{v_j})|i,j \leq B\}$
\end{algorithmic}
\end{algorithm}

This creates controlled mismatches at probability $p_{mix}$ by replacing $v_j$ with shuffled counterparts $\hat{v_k}$, forcing the model to develop robust audio understanding without over-relying on visual context.

The complete model optimizes standard captioning loss $\mathcal{L}_{CE}$ combined with regularization terms for gate stability and semantic consistency across perturbed examples. Our ablation studies demonstrate how these components synergistically improve performance under diverse mismatch conditions.

\section{Experiments}

\begin{table*}[t!]
\centering
\caption{Performance on the AudioCaps test set with the context of long video inputs (18 frames). \textbf{Bold} represents the best performance for each metric, and Params denotes trainable parameters.}
\label{tab:main}
\begin{tabular}{lc *{9}{S[table-format=1.3]}}
\toprule
\textbf{Model}  & \textbf{Params}  & \textbf{{Bleu$_1$}} & \textbf{{Bleu$_2$}} & \textbf{{Bleu$_3$}} & \textbf{{Bleu$_4$}} & \textbf{{METEOR}} & \textbf{{ROUGE$_L$}} & \textbf{{CIDEr}} & \textbf{{SPICE}} & \textbf{{SPIDEr}} \\
\midrule
Audio Only & 153 M    & 0.645      & 0.476      & 0.337      & 0.231      & 0.225    & 0.474       & 0.635   & 0.162   & 0.398 \\
AVCap   & 233 M       & 0.666      & 0.491      & 0.349      & 0.240      & \textbf{0.230}    & \textbf{0.479}       & 0.646   & 0.170   & 0.408  \\
\textbf{Proposed}   & 235 M     & \textbf{0.670}      & \textbf{0.500}      & \textbf{0.361}      & \textbf{0.258}      & 0.228    & 0.474       & 0.649   & 0.170 & 0.410   \\
\textbf{Proposed w/o SMS}  & 235 M   & 0.662      & 0.492      & 0.355      & 0.256      & 0.226    & 0.473       & \textbf{0.650}   & \textbf{0.172} & \textbf{0.411}   \\
\bottomrule
\end{tabular}
\end{table*}

\subsection{Experimental settings}

\subsubsection{Dataset and metrics}
We conduct experiments on the AudioCaps dataset~\cite{kim2019audiocaps}, the most extensive audio captioning corpus containing 51,308 YouTube video clips from AudioSet~\cite{gemmeke2017audioset} with human-annotated captions. Since some links are no longer accessible, we obtained 43,941 out of 49,838 audio-video clips for the train set, 447 out of 495 for the validation set and 866 out of 975 for the test set. Each clip contains 10-second audio paired with video frames. For the training set, each clip has one textual caption, while for the test set, each clip has five textual captions.

We adopt five conventional captioning metrics: BLEU$_n$(B$_n$, $n \in [1,\dots,4]$) \cite{papineni2002bleu}, METEOR (M) \cite{banerjee2005meteor}, ROUGE$_L$ (R) \cite{lin2004rouge}, CIDEr (C) \cite{vedantam2015cider}, and SPICE (S) \cite{anderson2016spice}, along with the SPIDEr \cite{liu2017spider} metric that combines CIDEr and SPICE. All metrics are calculated using the MS-COCO \cite{lin2014coco} evaluation toolkit. 

\subsubsection{Baselines}
Our experimental framework is established within the context of long video inputs (18 frames). Due to the unavailability of the multimodal model weights and associated training code from VACT, we exclusively present the experimental results of AVCap as our baseline. All models were trained on the training set using 18-frame video sequences as visual cues. The "Audio Only" configuration represents the AVCap model without visual information, maintaining identical architecture to other systems, and serves as the experimental lower bound.

\subsubsection{Implementation details}

Our implementation builds upon the AVCap architecture. For video processing, we uniformly sample 18 frames (2 FPS × 9s) and split each frame into 16×16 patches. We extract 128-dimensional log Mel-filterbank (FBank) features from an audio clip using a 25ms Hamming window with a 10ms stride and split them into 16×16 patches to match the video feature structure. Both audio and visual features are independently processed through modality-specific 11-layer ViTs with 768-dimensional hidden states and 12 attention heads.

\subsection{Main results}

The proposed method demonstrates competitive performance across all metrics on the AudioCaps test set. As shown in Table \ref{tab:main}. Our data augmentation variant achieves state-of-the-art results in all BLEU metrics, showing the effectiveness of data augmentation for n-gram-based evaluation. However, the base Proposed model without augmentation obtains the best scores in three semantic-aware metrics: CIDEr (0.650), SPICE (0.172), and SPIDEr (0.411), suggesting better semantic alignment despite slightly lower n-gram matches.


AVCap maintains superiority in METEOR (0.230) and ROUGE-L (0.479), indicating particular strengths in recall-oriented matching and synonym recognition. The audio-only baseline shows expected limitations across all metrics, confirming the value of multimodal approaches. A further analysis of the results can be found in Section \ref{sec:ablation}.

\subsection{Effect of data mismatch}

\begin{table}[tb]
\centering
\caption{Study on training-phase audiovisual shuffle probability (Prob), where Prob controls the likelihood of randomly re-pairing audio and video streams within each mini-batch to simulate mismatch scenarios during training. All models are evaluated on the original unshuffled test set (0\% shuffle).}
\label{tab:prob}
\scalebox{0.96}{
    \begin{tabular}{l *{6}{S[table-format=1.3]}}
    \toprule
    \textbf{Prob}         & \textbf{{B$_1$}} & \textbf{{B$_4$}} & \textbf{{M}} & \textbf{{R}} & \textbf{{C}} & \textbf{{S}} \\
    \midrule
    0\%     & 0.662      & 0.256      & 0.226    & 0.473       & \textbf{0.650}   & \textbf{0.172}  \\
    5\%     & \textbf{0.670}      & \textbf{0.258}      & \textbf{0.228}    &\textbf{ 0.474}       & 0.649   & 0.170    \\
    50\%    &0.656      &0.242      &0.221      &0.469      &0.631       &0.170     \\
    100\%   &0.664      &0.236      &0.223      &0.470      &0.622       &0.166     \\
    \bottomrule
    \end{tabular}
}
\end{table}

\begin{table*}[t]
\centering
\caption{Complete performance comparison under different shuffle probabilities of testset.}
\label{tab:exp_mismatch}
\begin{tabular}{ ll *{8}{c} }
\toprule
\textbf{Method} & \textbf{Probability} & \textbf{Bleu$_1$} & \textbf{Bleu$_2$} & \textbf{Bleu$_3$} & \textbf{Bleu$_4$} & \textbf{METEOR} & \textbf{ROUGE$_L$} & \textbf{CIDEr} & \textbf{SPICE} \\
\midrule
\multirow{4}{*}{AVCap}
& 0\%   & 0.666 & 0.491 & 0.349 & 0.240 & 0.230 & 0.479 & 0.646 & 0.170 \\
& 5\%   & 0.663 & 0.487 & 0.342 & 0.232 & 0.223 & 0.474 & 0.651 & 0.158 \\
& 50\%  & 0.664 & 0.487 & 0.341 & 0.231 & 0.223 & 0.472 & 0.642 & 0.158 \\
& 100\% & 0.657 & 0.481 & 0.338 & 0.229 & 0.223 & 0.470 & 0.636 & 0.157 \\
\midrule
\multirow{4}{*}{Proposed}
& 0\%   & 0.670 & 0.500 & 0.361 & 0.258 & 0.228 & 0.474 & 0.649 & 0.170 \\
& 5\%   & 0.669 & 0.498 & 0.359 & 0.255 & 0.227 & 0.473 & 0.649 & 0.169 \\
& 50\%  & 0.667 & 0.498 & 0.359 & 0.255 & 0.227 & 0.471 & 0.642 & 0.168 \\
& 100\% & 0.667 & 0.493 & 0.357 & 0.258 & 0.225 & 0.467 & 0.627 & 0.168 \\
\bottomrule
\end{tabular}
\end{table*}

\label{sec:ablation}

\begin{table}[t]
\centering
\caption{Ablation study on key components: "w/o S" denotes removing Stochastic Modality Shuffling, "w/o E" indicates disabling entropy gating, "w/o S\&E" removes both components, and "w/o G" represents baseline without gating mechanisms.}
\label{tab:ablation}
    \scalebox{0.93}{
        \begin{tabular}{l *{6}{S[table-format=1.3]}}
        \toprule
        \textbf{Method}         & \textbf{{B$_1$}} & \textbf{{B$_4$}} & \textbf{{M}} & \textbf{{R}} & \textbf{{C}} & \textbf{{S}} \\
        \midrule
        Proposed   & \textbf{0.670}      & \textbf{0.258}      & \textbf{0.228}    & 0.474       & 0.649   & 0.170    \\
        w/o S      & 0.662    & 0.256    & 0.226    & 0.473      & \textbf{0.650}   & 0.172 \\
        w/o E      & 0.660    & 0.253    & 0.227    & \textbf{0.475}      & 0.640   & \textbf{0.175} \\
        w/o S\&E   & 0.664    & 0.245    & 0.224    & 0.469      & 0.644      & 0.169    \\
        w/o G      & 0.654    & 0.248    & 0.226    & 0.475      & 0.643      & 0.168    \\
        \bottomrule
        \end{tabular}
    }
\end{table}

\subsubsection{Mismatching during training}
To validate the impact of audiovisual mismatch on system performance, we conducted model training with varying probability settings for SMS. The quantitative results are presented in Table~\ref{tab:prob}. 
We observed that when applying a shuffling probability of 5\% to simulate real-world mismatch conditions through perturbed audiovisual pairings, the model demonstrated improved recall metrics at the expense of reduced semantic accuracy (as measured by CIDEr scores). More extreme mismatch scenarios with higher shuffling probabilities adversely affected all evaluation metrics, suggesting that excessive artificial mismatching during training degrades overall captioning quality. This empirical evidence highlights the importance of maintaining an appropriate balance between mismatch simulation and the preservation of genuine audiovisual correspondence during the augmentation process.

\subsubsection{Mismatching during prediction}
Table \ref{tab:exp_mismatch} demonstrates the effectiveness of our proposed SMS through systematic evaluation under varying degrees of audiovisual mismatch. 

First, our method exhibits superior robustness to modality mismatch compared to the baseline AVCap. Across seven out of eight evaluation metrics (excluding CIDEr), the proposed approach shows strictly smaller performance degradation than the baseline when test-time shuffle probability escalates from 0\% to 100\%. This resilience is particularly evident in semantic metrics like SPICE, where our method preserves 98.8\% of the original performance compared to AVCap's 92.4\% retention.

Notably, our method trained with 5\% shuffle probability generalizes well to extreme mismatch conditions (100\% shuffle), suggesting that even limited exposure to artificial mismatch during training enables effective generalization. The controlled randomization in SMS acts as a regularizer, preventing over-reliance on visual modality while preserving beneficial cross-modal interactions.


\subsection{Ablation studies}
To validate the effectiveness of key components in our framework, we conduct comprehensive ablation studies under five experimental settings and quantitative results are presented in Table~\ref{tab:ablation}: (1) the full proposed model, (2) removing the SMS strategy, (3) disabling the attention entropy gating mechanism, (4) removing both SMS and entropy gating, and (5) baseline cross-attention fusion without any mismatch handling mechanisms. 

Three key observations emerge from the ablation studies. First, the model employing only traditional gating mechanisms (w/o S\&E) demonstrates performance comparable to that of directly utilizing cross-attention for modality fusion (w/o G), thereby substantiating the efficacy of our proposed gating strategy. 
Secondly, although w/o S or w/o E leads to improvements in semantic accuracy, the simultaneous elimination of both components w/o S\&E results in a decline in semantic precision. This observation underscores the complementary nature of our two innovations in addressing audiovisual incongruities.
Finally, the \textit{baseline cross-attention} (w/o G) exhibits a decline across all metrics. This validates our hypothesis that naive cross-modal fusion leads to error propagation when audiovisual mismatch occurs.



\subsection{Inference time comparison}

We conduct a comprehensive inference speed analysis to evaluate the computational efficiency of different approaches under varying video input lengths. The experiment employs 1024-frame audio clips paired with video inputs sampled at two frames per second (fps), with frame quantities set to {0, 1, 4, 8, 18} to simulate different temporal contexts. Each configuration undergoes five warm-up iterations before performing 10 timed inference runs, with the final results averaged for stability. Our proposed method demonstrates superior computational efficiency (2.369-2.733 seconds) compared to AVCap (3.416-15.029 seconds), mainly showing more significant advantages as video context expands. The audio-only baseline, operating without visual inputs, maintains a consistent inference time of 2.351 seconds. This systematic evaluation reveals that our architecture achieves significant speed improvements while handling multimodal inputs, especially in long-sequence processing scenarios.

\begin{figure}
    \centering
    \includegraphics[width=0.9\linewidth]{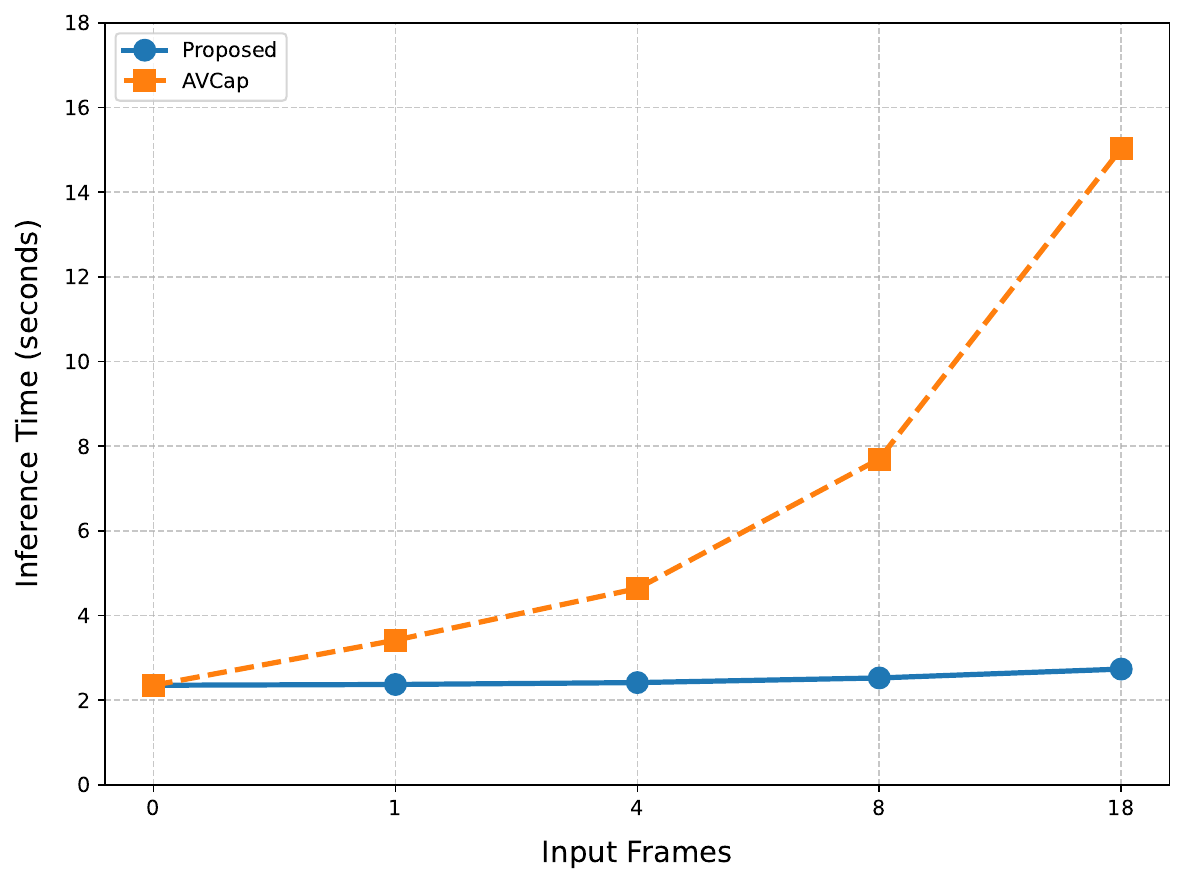}
    \caption{Inference time comparison (seconds) across different length of input video frames. 0 Input Frames denotes a baseline with no video modal inputs.}
    \vspace{-10pt}
    \label{fig:speed}
\end{figure}

\section{Conclusion}

In this paper, we present a novel framework to address audiovisual mismatch in video-assisted audio captioning through adaptive gating and mismatch-aware training. By leveraging cross-attention entropy as a self-supervised gate, our model dynamically suppresses irrelevant visual features when audio and video modalities are misaligned while preserving beneficial cross-modal cues in matched scenarios. Complemented by a batch-wise audiovisual shuffling strategy that simulates real-world mismatch noise, the approach achieves robust performance across both aligned and misaligned conditions. Experiments on standard benchmarks demonstrate state-of-the-art results, with ablation studies confirming the dual contributions of our gating mechanism and data augmentation. The solution requires no additional supervision or architectural complexity, offering practical value for real-world deployment. Future work will extend this paradigm to other multimodal tasks vulnerable to alignment noise.

\newpage


\bibliographystyle{IEEEtran}
\bibliography{mybib}

\end{document}